\documentclass{article}

\usepackage{gensymb}
\usepackage{PRIMEarxiv}

\usepackage[utf8]{inputenc} 
\usepackage[T1]{fontenc}    
\usepackage{hyperref}       
\usepackage{url}            
\usepackage{booktabs}       
\usepackage{amsfonts}       
\usepackage{nicefrac}       
\usepackage{microtype}      
\usepackage{lipsum}
\usepackage{fancyhdr}       
\usepackage{graphicx}       
\graphicspath{{media/}}     

\title{3-D poling and drive mechanism for high-speed PZT-on-SOI Electro-Optic modulator using remote Pt buffered growth}

\author{
  Suraj, Shankar Kumar Selvaraja \\
  Center for Nanoscience and Engineering \\
  Indian Institute of Science, Bangalore 560012, India \\
  \texttt{\{Suraj,Shankar Kumar Selvaraja\} suraj2@iisc.ac.in, shankarks@iisc.ac.in} \\
}

\begin{document}
\maketitle

\vspace{-2em}
\begin{abstract}
In this work, we have demonstrated a novel method to increase the electro-optic interaction in an intensity modulator at the C-band by optimizing the growth methodology of PZT with the metal (Ti/Pt) as a base material and the PZT poling architecture. Here, we have used a patterned Pt layer for PZT deposition instead of a buffer layer. By optimizing the PZT growth process, we have been able to do poling of the fabricated PZT film in an arbitrary direction as well as have achieved an enhanced electro-optic interaction, leading to a DC spectrum shift of 304 pm/V and a V$_\pi$L$_\pi$ value of 0.6 V-cm on a Si-based MZI. For an electro-optic modulator, we are reporting the best values of DC spectrum shift and V$_\pi$L$_\pi$ using perovskite as an active material.  The high-speed measurement has yielded a tool-limited bandwidth of > 12GHz. The extrapolated bandwidth calculated using the slope of the modulation depth is 45 GHz. We also show via simulation an optimized gap of $\approx$ 4.5 $\mu$m and a PZT thickness of $\approx$ 1 $\mu$m that gives us a less than 1 V-dB.
\end{abstract}


\section{Introduction}
\vspace{-0.6em}
A typical optical network consists of a laser source, an optical interconnect, and a modulation mechanism that can modulate the source optical parameters such as phase, intensity, or amplitude according to the information to be transmitted via the optical network. Lithium Niobate (LNO)\cite{wooten2000review,boes2018status,nelan2022ultra} has been a popular material used for intensity modulators due to the presence of pockels coefficient. Recent works on other perovskite materials such as PZT\cite{alexander2018nanophotonic,singh2021sputter,ban2022low} and BTO\cite{abel2016hybrid,ortmann2019ultra} have shown the potential to replace LNO based modulator in thin films due to pockel’s coefficient much larger than that of LNO. PZT (Lead Zirconium Titanate) has proved to be a promising material due to its ability to be integrated with Silicon with ease\cite{wang2022high,smith2012pzt,toyama1994characterization}. Recent reports on PZT-based modulators shown on Si or SiN platforms have mostly focused on the fabrication and integration aspect and not the effect of the crystallinity of the PZT film on the performance of the electro-optic modulator. Also, the effect of electrode configuration on the quality of polling has not been dealt with. Poling is one of the most critical steps in achieving a high-efficiency electro-optic modulator. Poling helps to align the maximum Pockels coefficient tensor in the direction of the applied electric field to achieve maximum DC shift of the optical spectrum and reduce the V$_\pi$L$_\pi$  parameter, both of which is a figure of merit in qualifying the electro-optic modulator. There have been reports on poling mechanism on bulk and thin film PZT\cite{kazansky1997electric,sakata1996sputtered,almusallam2014development}. Most of these reports have done a 1-dimensional poling. In our work, we have combined and optimized the film growth process with the poling mechanism to come at a novel method to align the domain of the film in an arbitrary direction to obtain maximum electro-optic interaction. This novel method relaxes the constraint on the growth mechanism to obtain an epitaxial film in the plane of electric field application. This reduces the V$_\pi$L$_\pi$ by increasing the overlap between the electric and optic field. Another design-based novelty in our approach that further reduces V$_\pi$L$_\pi$  is the absence of a buffer layer on top of the waveguide that further increases the electro-optic interaction.
In this work, we have, for the first time, used a metal (Ti/Pt) layer to fabricate an electro-optic modulator at C-band using PZT as an active material. We use the metal layer as a seed for PZT crystallization and as a bottom layer to perform the device polling. We have used an innovative interdigitated electrode configuration as a top and bottom electrode with PZT as an intermediate layer above Si MZI. We were able to achieve a DC shift of 228 pm/V with the proposed 3-D polling mechanism as compared to 3-6 pm/V achieved with co-planar polling. A V$_\pi$L$_\pi$ of 0.6 V-cm was achieved with the interdigitated electrode configuration. The reported value of DC shift and V$_\pi$L$_\pi$  have been the best-reported value using PZT or BTO (Barium Titanate) so far. High-speed measurements were done on the devices. A modulation bandwidth limited by tool capacity was observed to be >12 GHz.    

\section{PZT film growth and 3-D poling mechanism}

The effect of an electric field on the refractive index of a material can be given by eq.\ref{eq:index ellipsoid} and eq.\ref{eq:pockel}.

\begin{equation}
    \frac{x^2}{n{_1}{^2}}+\frac{y^2}{n{_2}{^2}}+\frac{z^2}{n{_3}{^2}}+\frac{2yz}{n{_4}{^2}}+\frac{2xz}{n{_5}{^2}}+\frac{2xy}{n{_6}{^2}}=1
    \label{eq:index ellipsoid}
\end{equation}

\begin{equation}
    \Delta\left(\frac{1}{n^2}\right)=\sum r{_{ijk}}E{^0}{_k}
    \label{eq:pockel}
\end{equation}

Taking PZT as a P4mm symmetry crystal lattice similar to BTO the modified index ellipsoid is given by eq.\ref{eq:modified pockel}.

\begin{equation}
    \left(\frac{1}{n{_o}{^2}} + r{_{13}}E{_z}\right)x{^2} + 
    \left(\frac{1}{n{_o}{^2}} + r{_{13}}E{_z}\right)x{^2} + 
    \left(\frac{1}{n{_e}{^2}} + r{_{33}}E{_z}\right)z{^2} + (r{_{51}}E{_y})2yz + (r{_{51}}E{_x})2zx = 1
    \label{eq:modified pockel}
\end{equation}

\begin{figure}[t]
\centering\includegraphics[width=\linewidth]{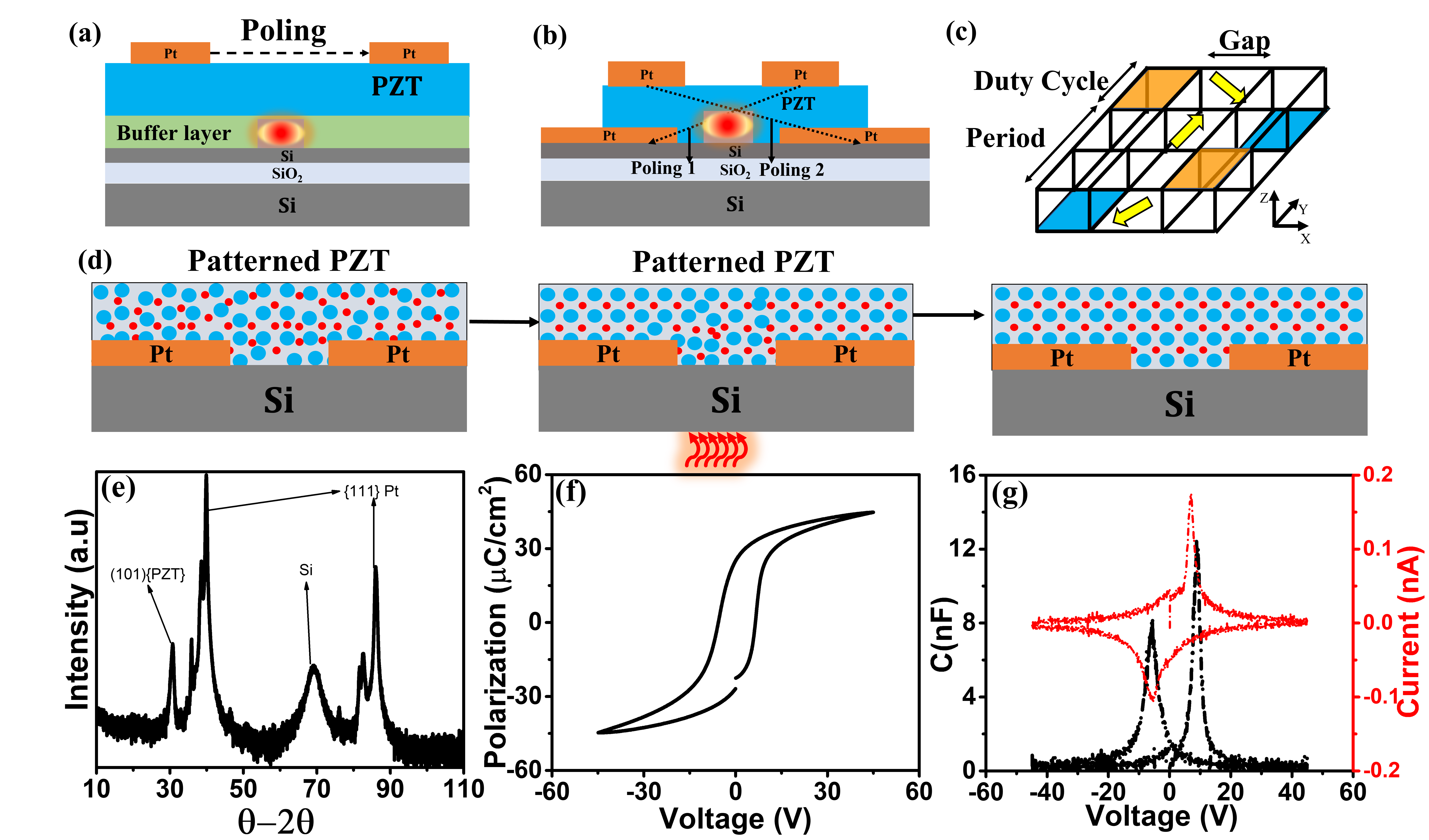}
\caption{Cross section schematic of EO modulator architecture (a) conventional,(b) proposed; (c) Proposed electrode configuration for 3D poling;(d) process flow showing deposition and annealing to obtain PZT thin film on patterned Pt; (e) $\theta$-2$\theta$ spectra of the deposited and annealed PZT film; electrical characterization of the deposited film showing (f) P-E loop hysteresis and (g) C-V and I-V characterization of the deposited PZT film.}
\label{fig:Material_process}
\end{figure}

As seen in Fig.\ref{fig:Material_process}(a) that the architecture conventionally used to fabricate EO modulator can be poled in 1-dimension using co-planar electrodes. As evident from eq.\ref{eq:pockel} that such architecture strictly requires an epitaxial film with the domain orientation of the film corresponding to an in-plane axis. As such, such architecture is inefficient for polycrystalline films or epitaxial films wherein the orientation of maximum Pockel's coefficient may not lie in the in-plane axis. Fig.\ref{fig:Material_process}(b) shows a proposed architecture that would enable us to access the bottom buffer electrode for poling with Fig.\ref{fig:Material_process}(c) showing the poling mechanism for such architecture. As is evident from Fig.\ref{fig:Material_process}(c) that the film can be poled in an arbitrary direction to obtain a peak value of pockels's. The corrugated nature of the electrodes makes the E$_x$ $\neq$ 0 enabling an EO effect in the zx direction too as shown in eq.\ref{eq:modified pockel}. The direction of the poling can be changed by varying the period($\Lambda$), electrode width (D), the gap between electrodes (g), and thickness of the PZT film (t$_{PZT}$). Fig.\ref{fig:Material_process}(d) shows the PZT film deposition and annealing methodology. The novelty in the process involves the use of a patterned Pt layer to deposit PZT and annealing to get a polycrystalline Pb(Zr$_{0.52}$Ti$_{0.48}$)O$_{3}$ with the Si in direct contact with the PZT thin film. The film was RF sputter deposited with an RF power of 70 W, argon flow rate of 30 sccm, post-annealing in air ambiance at 550\degree C at a ramp rate of 1.6\degree C/min. Fig.\ref{fig:Material_process}(e) shows the $\theta$-2$\theta$ XRD spectra of the deposited PZT film with (101) orientation with Fig.\ref{fig:Material_process}(f,g) showing the P-E loop, C-V and I-V characterization of the deposited and annealed film. The hysteresis in the P-E loop and the split peak in C-V confirms the ferroelectric behavior of the PZT film with a peak polarization of $\approx$50 $\mu$C/cm$^2$ and a peak capacitance of 12 nF, respectively.

\begin{figure}[t]
\centering\includegraphics[width=\linewidth]{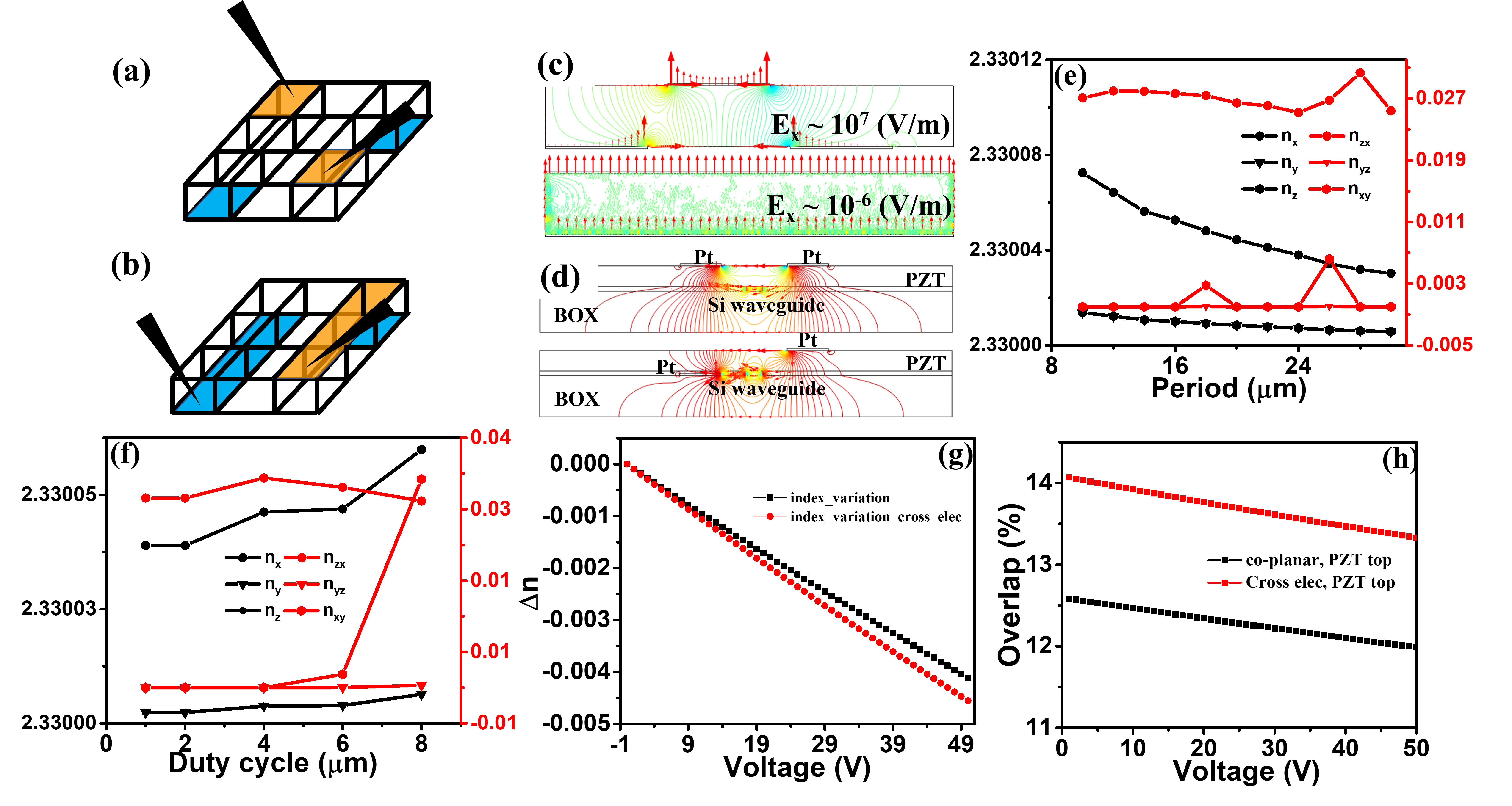}
\caption{Schematic of the electrode configuration for (a) corrugated electrode probing and (b) top-down electrode probing;(c)top view of the simulated electric field contours for a uniform co-planar electrode and corrugated co-planar electrode;(d) cross-section view of the simulated electric field contours for a co-planar electrode and top-bottom electrode; (c \& d) Ey and Ex field variation for 0 overlap; (e) Ex and Ey variation with overlap}
\label{fig:comsol_index_change}
\end{figure}

To investigate the effect of 3D poling, the structure in Fig.\ref{fig:Material_process}(c)  was decomposed into two structures shown in Fig.\ref{fig:comsol_index_change}(a\&b) and the electric field contours were compared with the conventional co-planar uniform electrode structure as seen in Fig.\ref{fig:comsol_index_change}(c\&d) in COMSOL 2D simulations.  As evident from Fig.\ref{fig:comsol_index_change}(c) that corrugated electrode configuration introduces E$_x$ while the increase in the overlap coefficient is observed in Fig.\ref{fig:comsol_index_change}(d). Fig.\ref{fig:comsol_index_change}(e-h) shows the COMSOL simulation to investigate the change in refractive index on the application of voltage with varying "$\Lambda$" and "D" of the electrode keeping the PZT thickness constant. As seen from Fig.\ref{fig:comsol_index_change}(e) shows the change in "n$_{zx}$" from a maximum value of 0.027 for a period of 10 $\mu$m to 0.020 for a $\Lambda$ of 24 $\mu$m with a "D" of 5 $\mu$m. Fig.\ref{fig:comsol_index_change}(f) shows the change in "n$_{zx}$" and "n$_{zx}$" with a maximum value of 0.035 for each at "D" of 4 $\mu$m and 8 $\mu$m, respectively. As evident in Fig.\ref{fig:comsol_index_change}(g) cross electrode probing is more efficient as compared to the co-planar electrode with maximum $\Delta$n of -0.005 and -0.004, respectively. Fig.\ref{fig:comsol_index_change}(h) shows the overlap of the electric field with the optic field with a maximum overlap of 14 \% for cross-electrode and 12.5 \% for co-planar electrode, respectively.

\begin{figure}[htbp]
\centering\includegraphics[width=\linewidth]{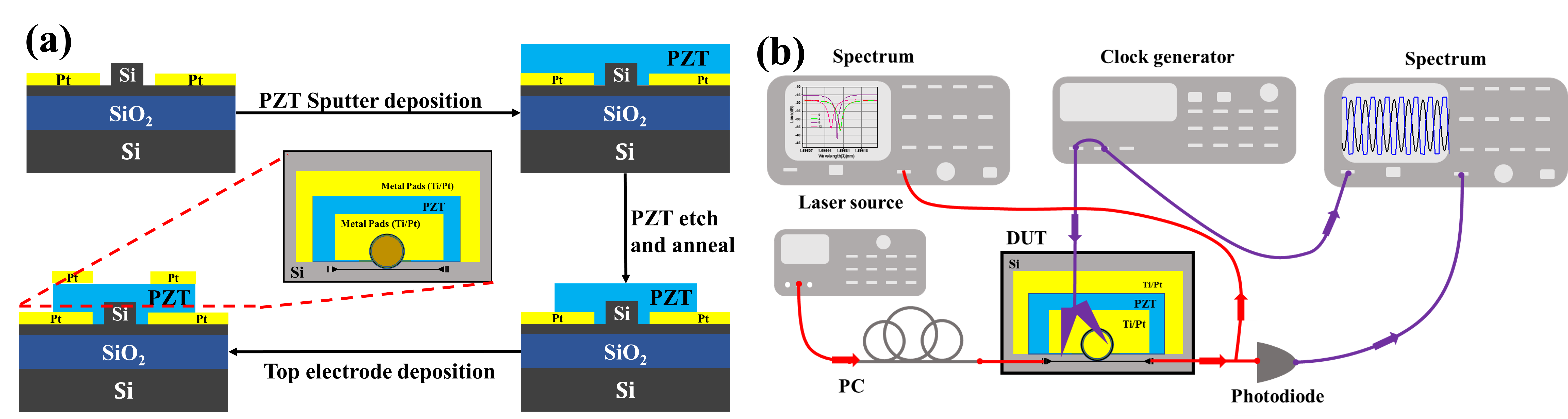}
\caption{(a)Process Flow to fabricate remote Pt buffered EO modulator;(b)measurement set-up used to characterize the fabricated modulator.}
\label{fig:Process_flow_paper}
\end{figure}

\subsection{Process Flow}

Fig. \ref{fig:Process_flow_paper}(a) shows the process flow to fabricate the electro-optic device. The fabrication starts with a passive Si MZI device fabricated on an SOI platform with a shallow etch of $\approx$80 nm. The bottom electrode of Ti/Pt (20/80 nm) is patterned using lift-off around the passive device with a post-annealing at 600\degree for 45 min. The lift-off is followed by PZT deposition and wet-etch patterning of PZT. The etching process exposes the bottom electrode that could be used for both poling and measurements. The deposition is performed at RF power of 70 W, argon flow rate of 30 sccm, post-annealing in air ambiance at 550\degree C at a ramp rate of 1.6\degree C/min. Top electrode of Ti/Pt (20/80 nm) is patterned on the PZT. Fig. \ref{fig:Process_flow_paper}(b) shows the measurement set-up used to characterize the fabricated modulator.

\begin{figure}[t]
\centering\includegraphics[width=\linewidth]{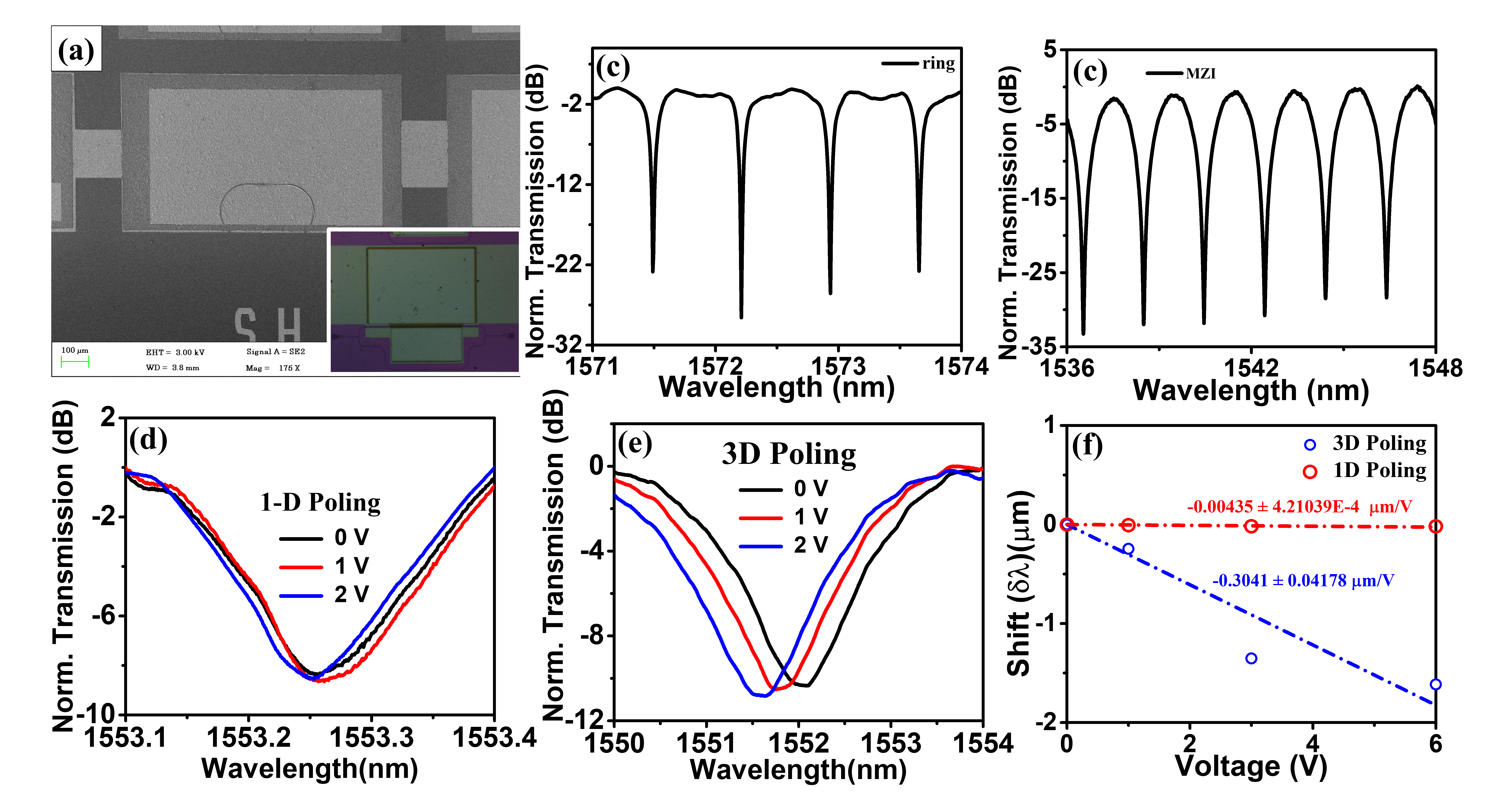}
\caption{(a) FESEM image of the fabricated racetrack ring resonator and optical microscopy of MZI; the Measured optical response of fabricated (b) racetrack ring resonator and (c) MZI; Post poling EO response of MZI using (d) co-planar electrode configuration and (e) corrugated electrode configuration; (f) comparison of 1D Poling and 3D Poling EO response.}
\label{fig:fab_DC_Simulation}
\end{figure}

\section{Measurements and Discussion}
\subsubsection*{DC electro-optic characterization}
Fig.\ref{fig:fab_DC_Simulation}(a) shows the fabricated ring resonator with the inset showing the optical microscopy of the fabricated MZI with an exposed bottom electrode for poling and measurement. Fig.\ref{fig:fab_DC_Simulation}(b,c)  shows the optical characterization of the fabricated racetrack ring and MZI resonator with the ring radius of \textbf{XX} and MZI $\Delta$l of \textbf{XX}. The poling mechanism used was the application of 70 V for a period of 45 min with 30 min cooling period using a co-planar electrode for 1D poling while for 3D poling the poling was performed using the top and bottom electrodes. As seen in Fig.\ref{fig:fab_DC_Simulation}(d,e) that the shift obtained for a 1D poling was $\approx$4 pm/V while for a 3D poling it was $\approx$ 300 pm/V. Fig.\ref{fig:fab_DC_Simulation}(f) shows the comparison of a 1D poling response with 3D poling response with a blue shift in both the cases and a slope of 4 pm/V and 304 pm/V respectively. The V$_\pi$L obtained for 3D poling is 0.6 V-cm. The obtained electro-optic response is the highest reported value for PZT EO modulator.

\subsection{High-speed measurements}

\begin{figure}[htbp]
\centering\includegraphics[width=\linewidth]{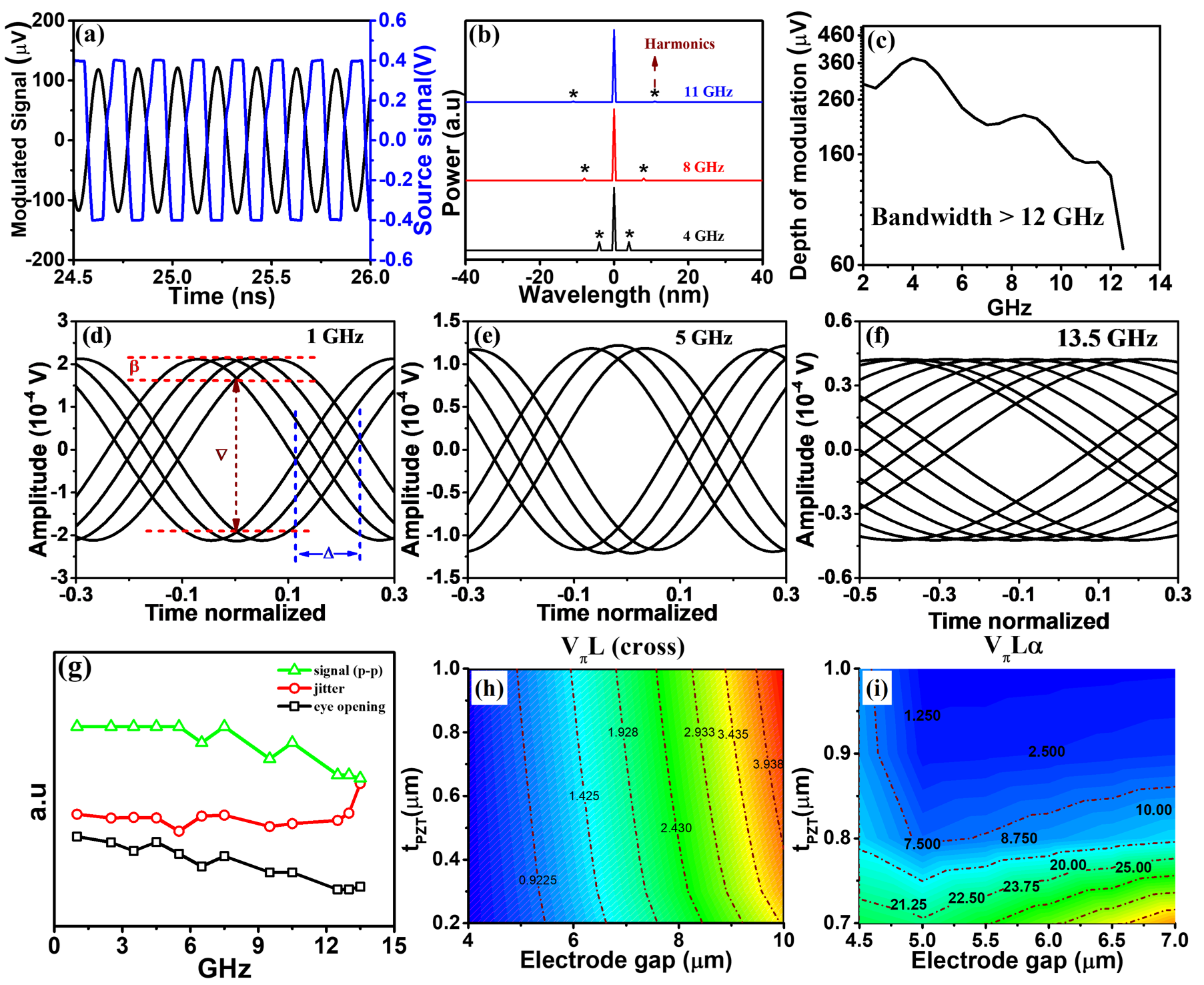}
\caption{(a) Time-domain curve showing the modulated optical output with varying electrical input; (b) frequency spectra of the modulated output with sidebands corresponding to modulating frequency;(c) Change in the depth of modulation with varying frequency of electrical input; Eye diagram corresponding to (d) 1GHz, (e) 5GHz and (f) 13.5 GHz; (g) comparison graph of the trend in jitter, modulating signal amplitude and eye-opening with varying modulating frequency; COMSOL simulation to optimize EO modulator (h)V$_{\pi}$L and (i)V$_{\pi}$L$\alpha$.}
\label{fig:EO_High_speed}
\end{figure}

Fig.\ref{fig:EO_High_speed}(a) shows the response of the EO modulator on the application of an electric field on a racetrack ring resonator. Fig.\ref{fig:EO_High_speed}(b) shows the frequency spectra of the modulated signal with the peak at 0 GHz corresponding to the carrier frequency and the sidebands corresponding to the modulating electrical signal that are characteristic of an intensity modulator. Using the peak-peak voltage as the depth of modulation, the bandwidth obtained is 12 GHz as is evident from Fig.\ref{fig:EO_High_speed}(c). The qualtity of modulation is determined by the eye opening($\nabla$), jitter($\Delta$) and the noise in the 0 or 1 level ($\beta$) as seen in Fig.\ref{fig:EO_High_speed}(d-f). As seen in Fig.\ref{fig:EO_High_speed}(g) at 12 GHz the $\nabla$ starts to decrease along with the modulating source amplitude (peak-peak). The slope of the source degradation is -0.031 V/GHz while the slope of eye opening is -0.0345 $\mu$V/GHz. The effective bandwidth, using $\beta$, taking the source amplitude degradation into account is theoretically calculated to be 45 GHz. The efficiency of the EO modulator can further be improved by varying the t$_{PZT}$ and electrode gap as seen in Fig.\ref{fig:EO_High_speed}(g,h) wherein t$_{PZT}$ of $\geq$ 0.9 $\mu$m and electrode gap $\geq$ than 5 $\mu$m gives a less than 1 (V dB). 
\begin{table}[]
\centering
\caption{Works on PZT based EO modulator }
\label{tab:Works on PZT based EO modulator}
\begin{tabular}
{|p{2cm}|p{3cm}|p{2cm}|p{2cm}|p{2cm}|}
\hline
Platform & DC EO response (pm/V) &  Bandwidth (GHz)& Deposition Method & V$_{\pi}$L (V-cm) \\
\hline   
SiN & 13 & 34 GHz  & sol-gel & 3.2 V-cm \\
\hline
Si {${\textbf{[this work]}}$} & 304 & 45 GHz  & Sputter	 & 0.6 V-cm \\
\hline
\end{tabular}
\end{table}

\section{Conclusion}

We have demonstrated the combined effect of the electrode design and the orientation of the PZT on the electro-optic interaction in an intensity modulator. By introducing Pt as a seed layer to pattern PZT on top of the Si MZI structure, we were able to circumvent the losses in electro-optic interaction due to the buffer layer which was an issue in the conventional electro-optic devices. By using the top and bottom interdigitated electrodes we fabricated a CMOS-compatible electro-optic modulator that can be poled in an arbitrary direction depending on the period, the duty cycle of the electrode, and the PZT thickness parameter. We were able to demonstrate DC MZI spectra shift of 228 pm/V using the 3D polling mechanism compared to 4 pm/V using co-planar polling. We achieved a V$_\pi$L$_\pi$ of 0.6 V-cm for the MZI intensity modulator with a tool-limited bandwidth of > 12 GHz. The performance of the modulator can further be improved by making the film epitaxial or by determining an optimum angle of poling for poly-crystalline PZT and hence making the design of the interdigitated electrode more deterministic.

\section*{Acknowledgments}
SKS thanks Professor Ramakrishna Rao chair fellowship.

\section*{Disclosures}
The authors declare no conflicts of interest.

\section*{Data availability}
Data underlying the results presented in this paper are not publicly available at this time but may be obtained from the authors upon reasonable request.

\bibliographystyle{unsrt}  
\bibliography{biblio_file}

\end{document}